\documentclass[conference]{IEEEtranNDSS}
\usepackage{versions}

\usepackage{xspace}

\usepackage{mfirstuc}
\usepackage[utf8]{inputenc}

\usepackage{listings}
\usepackage{mathtools}
\usepackage[frozencache=true,cachedir=.,newfloat=true]{minted}
\setminted{xleftmargin=1.5em,frame=lines,
framesep=2mm,
baselinestretch=1.2,
fontsize=\footnotesize,
linenos}

\usepackage{soul}

\usepackage{tcolorbox}
\usepackage{etoolbox}

\usepackage[acronym]{glossaries}
\usepackage{glossaries-extra}
\glsdisablehyper %

\setabbreviationstyle[acronym]{long-short}

\usepackage{multirow}
\usepackage{siunitx}
\usepackage{threeparttable}

\makeatletter
\newcommand\thefontsize[1]{{#1 The current font size is: \f@size pt\par}}
\makeatother

\usepackage[n,sets,adversary,operators,keys,notions,ff,primitives]{cryptocode}

\usepackage{xurl}
\usepackage{hyperref}
\hypersetup{breaklinks=true,colorlinks,citecolor=blue,linkcolor=blue,urlcolor=blue}
\usepackage{cleveref}
\usepackage{marginnote}
\usepackage[available,functional,reproduced]{ndssbadges}

\newcommand{\circnum}[1]{\raisebox{.5pt}{\textcircled{\raisebox{-.9pt} {#1}}}\xspace}

\newcommand{\inlinec}[1]{\texttt{#1}\xspace}

\newcommand{\design}{\gls{BliMe}\xspace}
\newcommand{\implboom}{\textsf{BliMe-BOOM}\xspace}

\newcommand{\blinded}{blinded\xspace}
\newcommand{\blindable}{blindable\xspace}

\newcommand{\blinds}{blinds\xspace}
\newcommand{\blindedness}{blindedness\xspace}
\newcommand{\unblind}{unblind\xspace}

\newcommand{\Blindable}{\xmakefirstuc{\blindable}}

\newcommand{\Blindedness}{\xmakefirstuc{\blindedness}}

\newcommand{\indicator}{tag\xspace}
\newcommand{\indicators}{\indicator{}s\xspace}

\newcommand{\srclink}{\url{https://drive.google.com/file/d/1tyMeQTwwP8GEPChRiuK4kKCcWkdSChdb/view?usp=sharing}}
\newcommand{\prooflink}[2]{\href{https://blinded-computation.github.io/blime-model/src/#1}{#2}}
\newcommand{\artifacturl}{\url{https://github.com/ssg-research/BliMe}}

\newcommand{\sreq}{\hyperlink{req.SR}{SR}\xspace}
\newcommand{\preq}{\hyperlink{req.PR}{PR}\xspace}
\newcommand{\creq}{\hyperlink{req.CR}{CR}\xspace}

\newif\ifgemfive\gemfivetrue

\newacronym{ABI}{ABI}{application binary interface}
\newacronym{API}{API}{application programming interface}
\newacronym{AWS}{AWS}{Amazon Web Services}
\newacronym{BliMe}{\textsf{BliMe}}{Blinded Memory}
\newacronym{BliMe-Ibex}{\textsf{BliMe-Ibex}}{Blinded Memory for Ibex}
\newacronym{BliMe-BOOM}{\textsf{BliMe-BOOM}}{Blinded Memory for BOOM}
\newacronym{CPU}{CPU}{central processing unit}
\newacronym{DIFT}{DIFT}{dynamic information flow tracking}
\newacronym{IFT}{IFT}{information flow tracking}
\newacronym{FHE}{FHE}{fully-homomorphic encryption}
\newacronym{FPGA}{FPGA}{field-programmable gate array}
\newacronym{IPC}{IPC}{inter-process communication}
\newacronym{IR}{IR}{intermediate representation}
\newacronym{ISA}{ISA}{instruction set architecture}
\newacronym{LUT}{LUT}{look-up table}
\newacronym{PIR}{PIR}{private information retrieval}
\newacronym{RAM}{RAM}{random access memory}
\newacronym{REE}{REE}{rich execution environment}
\newacronym{SoC}{SoC}{system-on-chip}
\newacronym{TA}{TA}{trusted application}
\newacronym{TCB}{TCB}{trusted code base}
\newacronym{TEE}{TEE}{trusted execution environment}
\newacronym{OS}{OS}{operating system}
\newacronym{DVFS}{DVFS}{dynamic voltage and frequency scaling}
\newacronym{OMP}{OMP}{oblivious memory partition}
\newacronym{OISA}{OISA}{oblivious instruction set architecture}
\newacronym{SEP}{SEP}{Secure Enclave Processor}
\newacronym{DRAM}{DRAM}{dynamic random-access memory}
\newacronym{HSM}{HSM}{hardware security module}
\newacronym{STT}{STT}{Speculative Taint Tracking}
\newacronym{PTE}{PTE}{page table entry}
\newacronym{CSP}{CSP}{cloud service provider}
\newacronym{SVFG}{SVFG}{static value-flow graph}
\newacronym{CFL}{CFL}{control-flow linearization}
\newacronym{DFL}{DFL}{data-flow linearization}
\newacronym{GPU}{GPU}{graphics processing unit}

\begin{document}

\author{
\IEEEauthorblockN{Hossam ElAtali}
\IEEEauthorblockA{\textit{University of Waterloo} \\ %
hossam.elatali@uwaterloo.ca}
\and
\IEEEauthorblockN{Lachlan J.\ Gunn}
\IEEEauthorblockA{\textit{Aalto University} \\ %
lachlan@gunn.ee}
\and
\IEEEauthorblockN{Hans Liljestrand}
\IEEEauthorblockA{\textit{University of Waterloo} \\ %
hans@liljestrand.dev}
\and
\IEEEauthorblockN{N.\ Asokan}
\IEEEauthorblockA{\textit{University of Waterloo} \\
\textit{Aalto University} \\
asokan@acm.org}
}

\IEEEoverridecommandlockouts
\makeatletter\def\@IEEEpubidpullup{6.5\baselineskip}\makeatother
\IEEEpubid{\parbox{\columnwidth}{
    Network and Distributed System Security (NDSS) Symposium 2024\\
    26 February - 1 March 2024, San Diego, CA, USA\\
    ISBN 1-891562-93-2\\
    https://dx.doi.org/10.14722/ndss.2024.24105\\
    www.ndss-symposium.org
}
\hspace{\columnsep}\makebox[\columnwidth]{}}

\title{{BliMe}: \\ Verifiably Secure Outsourced Computation \\ with Hardware-Enforced Taint Tracking}
\maketitle

\begin{abstract}
Outsourced computing is widely used today. However, current approaches for protecting client data in outsourced computing fall short: use of cryptographic techniques like fully-homomorphic encryption incurs substantial costs, whereas use of hardware-assisted trusted execution environments has been shown to be vulnerable to run-time and side-channel attacks.

We present \design, an architecture to realize efficient and secure outsourced computation. \design consists of a novel and minimal set of \gls{ISA} extensions implementing a taint-tracking policy to ensure the confidentiality of client data even in the presence of server vulnerabilities.
To secure outsourced computation, the \design extensions can be used together with an attestable, fixed-function \gls{HSM} and an encryption engine that provides atomic decrypt-and-taint and encrypt-and-untaint operations. %
Clients rely on remote attestation and key agreement with the \gls{HSM} to ensure that their data can be transferred securely to and from the encryption engine and will always be protected by \design's taint-tracking policy while at the server.

We provide an RTL implementation \implboom based on the BOOM RISC-V core. \implboom requires no reduction in clock frequency relative to unmodified BOOM, and has minimal 
power ($<\!1.5\%$) and FPGA resource ($\leq\!9.0\%$) overheads.
Various implementations of \design incur only moderate performance overhead (8--25\%). We also provide a machine-checked security proof of a simplified model \gls{ISA} with \design extensions.
\end{abstract}

\glsresetall
\section{Introduction}
Outsourced computation has become ubiquitous. 
While cost-effective, outsourced computation introduces confidentiality concerns because the clients’ sensitive data must be sent to the \gls{CSP}’s servers for processing by some (possibly third-party) application software. Malicious \glspl{CSP} or software can misuse client data. Furthermore, even if the \gls{CSP} is trusted, other malicious actors may compromise the sensitive data through run-time attacks or side-channel leakage.

One solution to this problem is to use cryptographic techniques such as \gls{FHE} which allow arbitrary computation on encrypted data. With \gls{FHE}, the server receives and processes only encrypted data. 
However, \gls{FHE} incurs very large performance overheads, orders of magnitude worse than processing the plaintext directly~\cite{viand21}. 
An alternative is to use hardware-assisted security mechanisms like \glspl{TEE} present in modern \gls{CPU} architectures, which can potentially ensure security while maintaining performance. A \gls{TEE} provides isolation for each client from the server \gls{OS} as well as from other clients. But bugs in server software can lead to run-time attacks. Furthermore, most \gls{TEE} implementations offer only limited resistance to side-channel attacks~\cite{intelSGXExplained2016,sanctum,demystifyingTrustzone}. Therefore, the need remains for an efficient and effective approach to protect outsourced computation even in the presence of software vulnerabilities and side channels.

Our goal is to address this need via minimal hardware changes. %
We design a minimal set of extensions for the RISC-V \gls{ISA} to preserve data confidentiality even in the presence of software vulnerabilities or side channels. The extensions and the accompanying attestation architecture, which together we call \design, allow a client to send conventionally encrypted data to a remote server, so that the \gls{CPU} can decrypt and process the data without allowing it or any data derived from it to be exfiltrated from the system. Computation results are returned only after encryption with the client's key.

We do this by having the \gls{CPU} enforce a taint-tracking policy preventing client data from being exported from the system. Prior work on hardware-enforced taint tracking~\cite{yu19} provide an untaint instruction to extract results, implicitly assuming that software invoking this instruction is trusted, making them vulnerable to run-time or speculative execution attacks~\cite{spectreDeclassified}. In contrast, \design uses a small attestable, fixed-function \gls{HSM} and an encryption engine to facilitate secure import and export of data between clients and servers; decryption on server-side always results in tainted plaintext. This allows \design to provide its security guarantees to \emph{multiple independent clients}, who \emph{do not need to trust server software}, including the \gls{OS}. Consequently, \design protects not just against side channels, but even against malware and run-time attacks that allow an attacker to execute arbitrary server software.
\ifdefined\IncSrcCode
\footnotetext[1]{Anonymized source code for review: \srclink.}
\fi
Our contributions are:
\begin{itemize}
    \item \design, an architecture with a set of taint tracking \gls{ISA} extensions for preventing exfiltration of sensitive data %
    (\Cref{sec:design}).
    \item \implboom, an RTL implementation of the \design \gls{ISA} extensions, incorporated into the speculative out-of-order RISC-V core BOOM (\Cref{sec:implementation}).
    \item A machine-checked proof of the dataflow security of the \design %
    taint-tracking policy %
    applied to a simplified model \gls{ISA} (\Cref{sec:securityEval}).
    \item A performance evaluation of \implboom 
    showing minimal run-time, power and area overheads (\Cref{sec:perfEval}).
\end{itemize}

\section{Background}

\subsection{Trusted Execution Environments}
\label{sec:bg-tee}

\Glspl{TEE} isolate \glspl{TA}---programs within \glspl{TEE}---from software outside the \gls{TEE} as well as from other \glspl{TA}. %
In addition to \gls{TA} isolation, \Glspl{TEE} also provide \emph{remote attestation} to assure clients that the code and configuration of server-side components are what they expect. \Glspl{TEE} are already present in several x86, Arm and RISC-V \glspl{CPU}, and are available to clients through \glspl{CSP} such as Amazon, Google, and Microsoft.

\Glspl{TEE} contain a root of trust for attestation, typically in the form of a unique attestation key embedded in the hardware at the time of manufacture. The remote attestation protocol first authenticates the hardware by having the server prove that its \gls{TEE} possesses this key, which is certified by the device manufacturer. After authentication, remote attestation is provided by ``measuring'' the system: the code, configuration, and state of the system are checked by the \gls{TEE} to assure the client that they are as expected.

Despite the claimed isolation guarantees, \Glspl{TEE} still suffer from certain vulnerabilities. Isolation only prevents malicious processes from naively accessing a \gls{TA}'s data through direct memory access. Remote attestation and code attestation only assure the clients that the system is set up as expected and that the code used to process the data is unmodified. Software bugs, present even in attested code, can lead to run-time attacks that circumvent client isolation, giving adversaries direct access to sensitive data. Despite many defenses, memory vulnerabilities and run-time attacks are still pervasive.

\subsection{Side-channel leakage}
\label{sec:side-channels}

Even if the software is bug-free, side-channel leakage can occur due to vulnerabilities or data-dependent behavior in the underlying hardware. Side channels are observable outputs of the system that are not part of the system's intended outputs. Prominent examples of \gls{CPU} side channels are execution time, memory access patterns, observable microarchitectural state (such as the state of shared caches, branch predictors and performance counters), voltage and electromagnetic radiation~\cite{paccagnellaLordRingSide2021,kocher99dpa,emAttack1,emAttack2,lippPLATYPUSSoftwarebasedPower2021,kimThermalBleedPracticalThermal2022}. Side-channel leakage can occur when an adversary is able to infer information about the sensitive data by observing the system while the data is processed. For example, if a conditional branch instruction depends on a sensitive value and the execution time of each branch is different, an adversary can infer some information about the sensitive value by monitoring the time it takes to complete the branch. Another example is when a sensitive value is used to index an array in memory; the memory access pattern, which is the sensitive address in this case, can change the observable state of a shared cache, or result in an observable request on the main memory bus.

Modern \glspl{CPU} employ a variety of performance optimizations in the form of speculation and out-of-order execution that amplify the leakage caused by data-dependent behavior~\cite{Lipp2018meltdown,Kocher2018spectre,schwarzZombieLoadCrossprivilegeboundaryData2019,vanbulck2018foreshadow,ragabRageMachineClear2021,vanbulck2018foreshadow,gotzfriedCacheAttacksIntel2017a,brasserSoftwareGrandExposure2017,leeInferringFinegrainedControl2017,chenSgxPectreStealingIntel2019,ryanHardwarebackedHeistExtracting2019,zhangTruSenseInformationLeakage2018}. This behavior is largely transparent to software developers, making it difficult to detect when benign-looking code can cause side-channel leakage.
\section{Problem Description}
\subsection{Usage scenario}\label{sec:usage}

The scenario we target is where several clients send data to a remote server for outsourced computation. Each client starts a session with the server, sends its data, and the server invokes some \emph{potentially untrusted, third-party} software (``server software'') to perform computation on the data (possibly in combination with the server's own data). Once computation completes, the results are sent to the client. Data import/export and computation can repeat multiple times per session. Data exchange between clients and servers is secured using authenticated encryption. Multiple clients may connect to the server simultaneously, leading to multiple parallel sessions.

Execution of server software that can leak any information about client data must be prohibited, even when attackers can run malware on the server, exploit vulnerabilities in the server software, or use side channels to extract data. In other words, sensitive client data must not flow to any observable output, nor to any other client.%

\subsection{Goals and objectives}

\newcommand{\goal}[2]{%
    \vspace{0.5em}
    \par\noindent%
    \newdimen\protocolstepwidth%
    \protocolstepwidth=\linewidth%
    \addtolength\protocolstepwidth{-2\fboxsep}%
    \fbox{\begin{minipage}[c]{\protocolstepwidth}\textbf{#1}: \emph{#2}\end{minipage}}%
    \vspace{0.5em}%
}

We now identify design requirements. The first relates to security.

\goal{\hypertarget{req.SR}{SR}---Confidentiality}{
    When a client provides sensitive input data to the server, no party other than that same client can infer anything about this data, other than its length.
}

Malware running on the server may attempt to gain access to the data, or the software processing the sensitive data may itself be malicious, but must not be allowed to reveal sensitive data outside the system, or to anyone other than the original client.

The next requirement relates to performance.

\goal{\hypertarget{req.PR}{PR}---Fast execution}{
    The design will not significantly reduce the performance of software accepted by the \gls{CPU}, compared to running the same software on a similar processor without such protections.
}

It is important to ensure that any solution does not excessively degrade performance. Elimination of side channels may prevent certain optimizations, resulting in some overhead, but some high-performance security-critical software has already been hand-written in assembly to eliminate side channels, and solutions that significantly reduce its performance may prove unsuitable in applications that make heavy use of such software.

The final requirement relates to backwards-compatibility.

\goal{\hypertarget{req.CR}{CR}---Backwards compatibility}{
    Software that does not leak sensitive data, by covert channels or otherwise, will run successfully.
}

A large portion of server software does not process sensitive data. It is important from a practical point that existing software can run on the new hardware: if this is not the case, then a new software stack will be needed, greatly limiting utility.

Moreover, there is also software that handles sensitive data but that already does so safely, such as side-channel-resistant cryptographic software.  It is equally desirable that this secure and well-tested software will continue to run on our hardware.

\section{Design}%
\label{sec:design}

\subsection{System overview}

\begin{figure}
    \centering
    \includegraphics[width=0.9\linewidth]{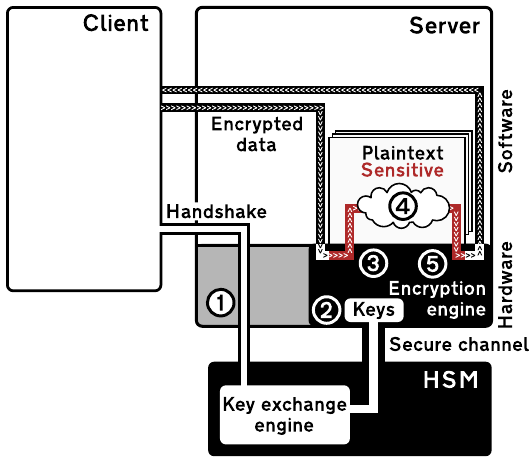}
    \caption{\design Overview. The client completes a cryptographic handshake with \gls{HSM}~\circnum{1}, which stores the resulting shared key in a protected register for use by the encryption engine~\circnum{2}. The client encrypts its secret data, which the server software decrypts with the aid of the encryption engine, while atomically marking the decrypted data as blinded~\circnum{3}. The server software can then perform the requested computation on the resulting \blinded data in a verifiably leakage-resistant manner~\circnum{4}, and encrypt the output using the encryption engine~\circnum{5} for the client.}
    \label{fig:system-overview}
\end{figure}

\Cref{fig:system-overview} shows an overview of \design. The server's software, including the \gls{OS}, runs on top of a \gls{CPU} that contains 1) the \design extensions, which enforce a taint-tracking policy on all software running on the \gls{CPU} (\Cref{sec:taint-tracking}), and 2) a \design encryption engine, used for data import and export (\Cref{sec:data-import,sec:data-export}). A \gls{HSM} is used to perform remote attestation, assuring the client that the server uses \design. The \gls{HSM} is a separate fixed-function component sharing few resources with the \gls{CPU}, reducing its exposure to side-channel attacks. The client knows how to verify the root of trust for attestation embedded within the \gls{HSM}, which contains a key exchange engine responsible for negotiating a session key between with the client. The \gls{HSM} provides this session key securely to the encryption engine without exposing it to the server software. %

\subsection{Adversary model}
We suppose that the server hardware, including the \design extensions, encryption engine, and the \gls{HSM}, is implemented correctly.

The adversary has control over all server software, including the \gls{OS}, and can make it behave as it sees fit, including making inferences based on side channel information such as memory access patterns and instruction traces.

Attacks against the hardware itself are currently out-of-scope; the client assumes that the attacker cannot use physical means to make the hardware act differently from its specification. We discuss this in \Cref{sec:discussion:other-attacks}. Side-channel attacks that require physical access (e.g., differential power analysis) are also out-of-scope.

\subsection{Protocol}\label{sec:design-protocol}

In this section, we outline the steps needed to perform safe outsourced computation using \design.

\subsubsection{Remote attestation and key agreement}\label{sec:remoteAttest}

Before sending any data to the server, a client first performs a handshake (\Cref{fig:system-overview}-\circnum{1}) with the \gls{HSM}, which consists of remote attestation and agreement on a session key. Remote attestation is provided by the \gls{HSM} to assure the client that the assumptions made in the adversary model hold. It attests two properties. First, the \gls{HSM} attests that it is genuine using the root-of-trust embedded within it at manufacture, as described in \Cref{sec:bg-tee}. Second, the \gls{HSM} attests to the client that it is embedded in server hardware incorporating \design.

At the end of remote attestation, the client and \gls{HSM} agree on a client-specific session key. The \gls{HSM} stores this key inside the encryption engine using a secure channel (\Cref{fig:system-overview}-\circnum{2}), along with a unique \blindedness \indicator identifying the client from which data was sourced; this is most simply implemented cryptographically using a sealing key shared by the two components. The encryption engine uses the session-specific key in two atomic functions that it exposes to the server software: \emph{data import} and \emph{data export}.%

\subsubsection{Data import}\label{sec:data-import}

Once the handshake is complete, the client locally encrypts its data using the session key, and sends the resulting ciphertext to the server. The server software calls the encryption engine's data import function on the ciphertext. The encryption engine then \emph{atomically} decrypts the ciphertext using the sealed session key, and taints the resulting plaintext by \emph{marking} it as \emph{``\blinded''} (\Cref{fig:system-overview}-\circnum{3}). This is done by setting a \emph{\blindedness} \indicator attached to the data in registers and memory. 
The \blindedness \indicator is an $n$-bit integer, which takes the value zero when its associated data is not \blinded. Other values indicate that the data is \blinded, and from which of $2^{n} - 1$ clients the data is sourced. In the minimal case where $n = 1$, this \indicator is a single bit that indicates whether or not the data is \blinded.

\begin{tcolorbox}
[width=\linewidth, colback=white!95!black]

Data import atomically decrypts data using the session key and marks it as \emph{\blinded}.

\end{tcolorbox}

\subsubsection{Safe computation}\label{sec:safe-comp}

Now, the server software can perform the requested computation on the \blinded plaintext data (\Cref{fig:system-overview}-\circnum{4}). The \design CPU extensions apply a taint-tracking policy, limiting the operations that can be done on \blinded data; this prevents the server software from directly exfiltrating the data or even leaking it through side channels. Any operations forbidden by the taint-tracking policy cause the server to fault. The policy ensures that the final results and any intermediate results derived from the data are also \blinded. More details are in \Cref{sec:taint-tracking}.

\subsubsection{Data export}\label{sec:data-export}

Once the computation is complete, the server software calls the encryption engine's data export function to atomically encrypt the \blinded results and mark the ciphertext as non-\blinded (\Cref{fig:system-overview}-\circnum{5}), which is done by zeroing the \blindedness \indicator. The server software can then send the ciphertext back to the client, who can decrypt it. The encryption engine ensures that \blinded data is only encrypted with the session key corresponding to its \blindedness \indicator. As the adversary controls the data to be encrypted, this encryption must be secure against adaptive chosen-plaintext attacks in order to prevent the adversary from using this ability to identify the ciphertexts corresponding to particular plaintexts.

\begin{tcolorbox}
[width=\linewidth, colback=white!95!black]

Data export atomically encrypts data using the session key and marks it as \emph{non-\blinded}.

\end{tcolorbox}

\subsection{Taint-tracking policy}\label{sec:taint-tracking}
We use taint tracking to \emph{prevent sensitive data from flowing to observable outputs}. First, we define which parts of the system can be tainted. We split the system state into two types:
\begin{itemize}
\item \emph{\Blindable} state consists of the values (not addresses) of lines in the cache, values in registers except the program counter, and values in main memory, as well as all busses and queues used to transfer these values. Each of these is extended with a \blindedness \indicator.
\item \emph{Visible} state consists of information that may be exposed outside the system, and must therefore never contain sensitive data. It includes \emph{all microarchitectural state that does not have a \blindedness \indicator associated with it}, e.g., the program counter, addresses of lines in the caches, branch predictor state, and performance counters. As the program counter encapsulates control flow, making it part of visible state forbids \blinded data from affecting control flow.
\end{itemize}

We then define the list of observable outputs we consider in \design: visible state, non-\blinded \blindable state, the addresses of memory operations sent to main memory, the execution time of an instruction or set of instructions, and fault signals. Note that once \blindable state becomes \blinded, it is no longer observable. We exclude outputs that require physical access to be observed, such as voltage and electromagnetic radiation.

\begin{table*}
    \centering
    \begin{threeparttable}[b]
        \begin{tabular}{c||c|c|c|c||c|c}
            \textbf{Instruction} & \multicolumn{4}{c||}{\centering \textbf{Tag (\O=unblinded, ?=``don't care'')}} & \textbf{Decision} & \textbf{Result tag} \\
            \hline
            \multirow{7}{2.5cm}{\centering \textbf{Arithmetic/Logic\tnote{\dag}}} & \multicolumn{4}{l||}{If \# of cycles depends on value of any blinded operand} & Fault & - \\
            \cline{2-5}
             & \multicolumn{4}{c||}{Otherwise:} & & \\
             & \multicolumn{2}{p{3.5cm}|}{\centering \textbf{Op1}} & \multicolumn{2}{c||}{\centering \textbf{Op2}} & & \\
            \cline{2-5}
             & \multicolumn{2}{p{3.5cm}|}{\centering \O} & \multicolumn{2}{c||}{\centering \O} & Propagate & \O \\
             & \multicolumn{2}{p{3.5cm}|}{\centering $a$} & \multicolumn{2}{c||}{\centering \O} & Propagate & $a$ \\
             & \multicolumn{2}{p{3.5cm}|}{\centering $a$} & \multicolumn{2}{c||}{\centering $a$} & Propagate & $a$ \\
             & \multicolumn{2}{p{3.5cm}|}{\centering $a$} & \multicolumn{2}{c||}{\centering $b$} & Fault & - \\
            \hline
            \multirow{4}{2.5cm}{\centering \textbf{Branching}} & \multicolumn{2}{p{3.5cm}|}{\centering \textbf{Addr}} & \multicolumn{2}{c||}{\textbf{Condition ops}} & & \\
            \cline{2-5}
             & \multicolumn{2}{p{3.5cm}|}{\centering \O} & \multicolumn{2}{c||}{\centering \O}       & Propagate & \O \\
             & \multicolumn{2}{p{3.5cm}|}{\centering $a$}    & \multicolumn{2}{c||}{\centering ?} & Fault & - \\
             & \multicolumn{2}{p{3.5cm}|}{\centering ?}    & \multicolumn{2}{c||}{\centering $a$}          & Fault & - \\
            \hline
            \multirow{4}{2.5cm}{\centering \textbf{Load}} & \multicolumn{2}{p{3.5cm}|}{\centering \textbf{Addr}} & \multicolumn{2}{c||}{\textbf{Data in memory}} & & \\
            \cline{2-5}
             & \multicolumn{2}{p{3.5cm}|}{\centering \O} & \multicolumn{2}{c||}{\centering \O}    & Propagate & \O \\
             & \multicolumn{2}{p{3.5cm}|}{\centering $a$}    & \multicolumn{2}{c||}{\centering ?}       & Fault & - \\
             & \multicolumn{2}{p{3.5cm}|}{\centering \O} & \multicolumn{2}{c||}{\centering $a$}       & Propagate & $a$ \\
            \hline
            \multirow{4}{2.5cm}{\centering \textbf{Store}} & \multicolumn{2}{p{3.5cm}|}{\centering \textbf{Addr}} & \multicolumn{2}{p{3.5cm}||}{\centering \textbf{Data in register}} & & \\
            \cline{2-5}
             & \multicolumn{2}{c|}{\O} & \multicolumn{2}{c||}{\O}       & Propagate                         & \O \\
             & \multicolumn{2}{c|}{$a$}    & \multicolumn{2}{c||}{?}          & Fault                        & - \\
             & \multicolumn{2}{c|}{\O} & \multicolumn{2}{c||}{$a$}          & Propagate                    & $a$ \\
        \end{tabular}
        \begin{tablenotes}
            \item [\dag] See \Cref{sec:secret-dep-faults} for a discussion on data-dependent faults.
        \end{tablenotes}
        
        \caption{BliMe taint-tracking policy rules for all instruction types. Tags $a$ and $b$ are arbitrary but different. Swapping them produces an equivalent result.}
        
        \label{tab:taint-tracking-2}
    \end{threeparttable}
\end{table*}

The taint-tracking policy is defined as follows (and as shown in \Cref{tab:taint-tracking-2}). An instruction with a \blinded input (i.e., which takes at least one input with a non-zero \blindedness \indicator) yields a \blinded value for each output that depends on a blinded input, with the same \blindedness \indicator as the input.
An instruction that receives \blinded data from multiple sources will raise a fault. If an instruction attempts to affect any observable output except non-\blinded \blindable state in a manner that depends on the value of a \blinded input, a fault is raised, since this can otherwise be used to exfiltrate sensitive data. This effectively means that the program cannot use a \blinded value as the address of a jump, branch or memory access, or use instructions whose completion time or fault status depends on a \blinded value. For example, \design prevents
\begin{itemize}
    \item \emph{Cache-timing side-channel attacks} by prohibiting \blinded values from being used as addresses for loads/stores, and
    \item \emph{Timing attacks} by forbidding \blinded-value-dependent control flow (prohibiting both altering PC based on \blinded values, and instructions of variable duration from using \blinded values). Measuring program execution time will reveal nothing about the \blinded values; the adversary will obtain the same information as if they ran the same program over a \blinded array of zeroes of the same length.
    \item \emph{Transient execution attacks} by preventing any \blinded values from being used in speculation decisions. Speculatively executed instructions can still use \blinded values (and maintain their \blindedness tags throughout this transient execution), but the result of any prediction \emph{decision} (e.g., whether a branch is taken, or what the next return address might be) cannot rely on \blinded values. This same concept applies to all data-dependent hardware optimizations; \blinded values cannot be used to \emph{decide} on whether or how an optimization is applied.
\end{itemize}

The \blindedness of any given value is not sensitive; this means that it is safe for the ISA to include instructions that query whether or not a given value is blinded.

\subsection{\design-compliant software}
\label{sec:compliant-software}
\design's restrictions on application software are the same as those required of anyone developing secure side-channel-resistant code (even without \design hardware modifications): they must adhere to constant-time coding principles~\cite{aumassonCryptocoding2022}, including data-oblivious control flow and memory access, and must not explicitly exfiltrate sensitive data outside the system. An example of such practices being used today can be found in the development of cryptographic libraries, the compatibility of which we show in \Cref{sec:compEval}.
Concretely, for software to be \design-compliant, it \underline{must not attempt} to:
\begin{enumerate}
    \item use \blinded data in any control-flow decisions,
    \item use \blinded data as the target address for any jump or branch instructions,
    \item use \blinded data as the address for any memory operations, including using \blinded data as an offset or index,
    \item use \blinded data as an operand for an instruction whose execution time depends on the value of that operand (e.g., variable-time division instruction),
    \item mix \blinded data belonging to separate security domains, i.e., with different \blindedness tags, and
    \item write \blinded data to any peripherals, e.g., displays, network devices, disk drives.
\end{enumerate}

Note that \design does \emph{not} require software developers to be aware of how the \gls{CPU} behaves speculatively. \design ensures that all \blinded data cannot be leaked by hardware optimizations, including speculation (\Cref{sec:taint-tracking}).

These requirements do not limit what can be computed.  A much more restrictive historical CPU, Zuse's Z3, lacks conditional branching and indirect addressing but can simulate any finite-tape Turing machine~\cite{rojas98}. \design allows conditional branching and indirect addressing on non-sensitive data as well as all of Z3's functionality, making it as powerful as but more efficient than Z3. Therefore, \design  can simulate any finite-tape Turing machine.

\section{Implementation}
\label{sec:implementation}

In this section, we first present the common architectural changes required to implement \design (\Cref{sec:implarch}), and then describe how these were applied to the out-of-order RISC-V BOOM core~\cite{boom} to obtain \implboom (\Cref{sec:boom}).

The implementation covers the hardware needed for safe computation (\Cref{sec:safe-comp}), including taint tracking and enforcement of the security policy (\Cref{sec:taint-tracking}). We do not include the \gls{HSM} in the implementation. Components with similar functionality already exist, such as Google's Titan-M chip~\cite{titanM} or Apple's \Gls{SEP}~\cite{SecureEnclave}; the HSM is configured to perform the attested handshake and provide correct client IDs, and the system integrator provides the secure channel between HSM and encryption engine in the form of a shared key embedded in each component.

\subsection{Architectural changes}
\label{sec:implarch}

The architectural changes needed to implement \design can be grouped into two main categories: taint tracking, and handling of policy violations.

\noindent\textbf{\underline{Taint-tracking}}
 is performed on registers and memory. Due to the load-store architecture of RISC-V, these can be handled separately.

\textbf{Registers and ALU operations.}
For each register, we maintain a \blindedness \indicator. Any ALU operation that reads from a \blinded register, \blinds its output register by default. However, this is a conservative approximation, and implementations can make exceptions in order to more accurately model instruction dataflows. For example, if a register is XORed with itself, the result is not \blinded because the result is always zero, irrespective of the input value. Similar exceptions can be used for other situations in which an instruction takes \blinded inputs but its output does not depend on said \blinded inputs.%

\textbf{Memory.}
\Blindedness is tracked in using a tagged-memory approach, with a \blindedness \indicator being attached to each physical address, in both main memory and in each cache. Whenever a value is stored into memory from a register, the memory bytes inherit the \blindedness of the register. The reverse holds for memory reads. 
To reduce overheads, multiple consecutive bytes in memory, forming a \emph{tag granule}, can share the same tag. For tag granule sizes larger than a single byte, we introduce an additional \design policy rule to prevent mixing \blinded data belonging to separate security domains: if a store instruction attempts a \emph{partial} write of \blinded data (e.g., a byte) with tag $a$ to a granule in memory (e.g., of size 8 bytes) with tag $b$, the instruction must fault.

\textbf{New Instructions.}
We introduce two new instructions, \inlinec{blnd} and \inlinec{rblnd}, that correspond to the data import and export operations, respectively (\Cref{sec:data-import,sec:data-export}). The instructions encrypt and decrypt data with a supplied key, writing the result into memory in un\blinded or \blinded form respectively.

\noindent\textbf{\underline{Handling violations}}
is needed in four situations:
\begin{enumerate}
    \item Attempting to write \blinded values to the \inlinec{PC} register: Jumps and conditional branches relying on \blinded registers, either as a jump destination address or as part of a conditional check, are forbidden.
    \item Attempting to use \blinded values in an instruction whose execution time depends on the \blinded input value.
    \item Attempting to read from or write to memory using a \blinded value as an address: This occurs when a load or store uses a \blinded register either as the address base or offset.
    \item Attempting to write \blinded data to an ``unblindable'' memory location. This allows a system to specify whether certain memory-mapped peripherals have access to \blinded data.
\end{enumerate}

Any of the above causes an illegal instruction fault.

We implemented the architectural changes in Spike~\cite{SpikeRISCVISA2022}, a C++ RISC-V ISA simulator, to enable quick testing of our extensions.

\subsection{\implboom}\label{sec:boom}

To demonstrate \design on a complex and realistic computation platform that relies heavily on speculation to achieve high performance, we implemented \design in RTL on the BOOM~\cite{boom} core and Chipyard~\cite{chipyard} \gls{SoC}. \design taint-propagation rules are also enforced during any speculative execution thereby preventing sensitive values from being leaked even by Spectre-type attacks~\cite{Kocher2018spectre}.

We implemented two variations of \design with different tag configurations: \implboom-1 with single-bit \blindedness \indicators and byte-level granularity, and \implboom-8 with eight-bit \blindedness \indicators and 8-byte (i.e., word-level) granularity.

The BOOM core is written in Chisel, which is a hardware construction language embedded in Scala~\cite{chisel}. Chisel defines common built-in data types, such as unsigned integers (UInt) and Boolean values (Bool), used to attach semantic meaning to bits in digital circuits. Chisel also allows user-defined data types to be created by composition and inheritance. Taking advantage of this feature, we create a \emph{Blinded} data type as a composition of a \blindedness \indicator and a variable-length vector of bits, and use this type throughout the design. As Scala (and, by extension, Chisel) is strongly-typed, this allows the compiler to ensure that \blinded values cannot be separated from their \blindedness \indicators except by code that is aware of the \emph{Blinded} type. It also allowed us to identify \emph{all locations in the code} where \blinded data can propagate and prevent any unintended leakage caused by hardware optimizations such as speculation.

\subsubsection{Registers}
Register files are modified to use the \emph{Blinded} type, thereby storing a \blindedness \indicator alongside the register contents.

\subsubsection{Main Memory \& Caches}
A region of memory is allocated for the storage of \blindedness \indicators, and made inaccessible to software.
We expand the L1 data and L2 caches to include \blindedness \indicators alongside each granule of data. The L1 instruction cache does not need \blindedness \indicators; \blinded values are not allowed to be executed, so  whenever a blinded value is read into the cache, it is replaced with a zero, and its `valid instruction' metadata bits are unset, making it inaccessible. The L2 cache includes logic to translate misses into \emph{two} operations to main memory: one for the actual data, and one for the \blindedness \indicators.

It is also possible to optimize the design for the specific case where the \blindedness \indicators are small relative to their associated data, with carefully-designed logic reducing the ratio of \blindedness \indicator requests to data requests, thereby reducing overhead. However, a limitation in Chipyard's TileLink implementation allows data to be read only at an 8-word granularity, yielding a 64-bit effective minimum tag size. Therefore, even for 8-bit tags, the memory controller ties up the bus for seven additional cycles to transmit unneeded words. This is not a limitation of \design, but of Chipyard. We forego implementing the extensive modifications required to the memory controller and instead investigate the effect of this optimization using gem5~\cite{gem5} in \Cref{sec:gem5}.

\subsubsection{Pipeline Stages}
In BOOM, instructions are decoded in the instruction decoder stage into micro-ops, which are stored in the reorder buffer until they are issued to an execution unit and committed. Execution units contain different functional units that can perform specific operations. We modified all functional units to either propagate \blindedness (e.g., addition of \blinded values), or fault when the operation is unsafe to perform with \blinded operands (e.g. memory address generation).

The functional unit responsible for memory address generation faults when blinded operands are used to generate a virtual address. Furthermore, if the page table walker finds a blinded \gls{PTE}, it zeroes it before storing it in the translation-lookaside buffer (TLB). This ensures that virtual-to-physical address translation by the load-store unit cannot be used to load from a blinded physical address. Any load or store attempting to use a blinded physical address directly will fault.

\subsubsection{Encryption engine}
The encryption engine, which performs atomic blind-and-decrypt and encrypt-and-unblind operations using the ChaCha20 stream cipher, is built as a RoCC accelerator~\cite[p.~3]{rocket} containing a pipelined ChaCha20 implementation with a 19-cycle latency. This allows this functionality to be accessed using a set of instruction opcodes reserved for accelerators, and allows it to be used by an unmodified C compiler using RISC-V intrinsics.

\subsubsection{Speculation}
Speculatively executed instructions are subject to the same checks as non-speculative instructions regarding blinded operands. Branch prediction can also never depend on blinded data. This is done by blocking all blinded data flows into the prediction logic. We 1) zero out any blinded instruction fetched into the instruction cache, 2) prevent any feedback from the execution units to the prediction logic regarding faulting branch decisions based on blinded data. By ensuring that both the speculation decision \emph{and} the speculative execution do not leak sensitive data, we block all speculative side channels. This is in a similar vein to \gls{STT}~\cite{STT} but, unlike \gls{STT}, we do not untaint after the instructions leave the speculative state and the security policy on the data remains in effect.

Note that, for \implboom, only non-constant-time operations, which always fault or are rolled back, affect speculation. Constant-time code, with no secret-data-dependent branching or memory accesses, speculates normally \emph{without overhead}. This is because BOOM does not have any other data-dependent speculation (e.g., data-dependent prefetchers). For microarchitectures with such speculation, it must be disabled (only) on blinded data.
\section{Security evaluation}
\label{sec:securityEval}

\subsection{Protocol}\label{sec:eval-protocol}
\begin{figure}
    \centering
    \includegraphics[width=\linewidth]{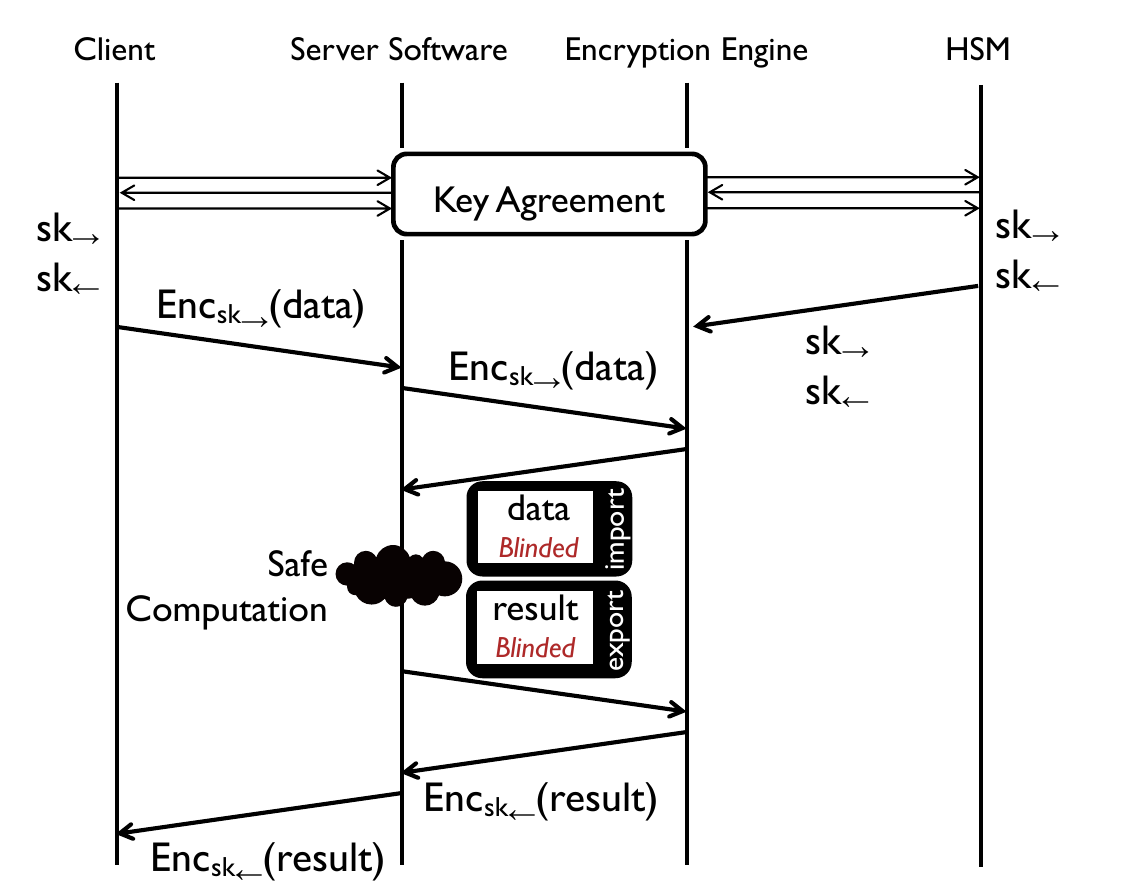}
    \caption{Protocol summary.  A key agreement protocol is used to obtain secret keys, shared by the client and \gls{HSM}.  The \gls{HSM} securely transports these keys to the encryption engine, which atomically decrypts and blinds the client data, performs safe computation on it, then atomically encrypts and blinds the client data before it is returned to the client.}
    \label{fig:protocol}
\end{figure}

Computation using \design takes the form of the protocol shown in \Cref{fig:protocol}. This is equivalent to the protocol described in \Cref{sec:design-protocol}. The attacker obtains only encrypted client data, the encryption of an arbitrary computation on the plaintext data, along with leakage from the computation by the server code, shown as a cloud in \Cref{fig:protocol}.

If the key agreement protocol and secure channel protocols are secure against active attackers, then the messages provide the attacker with no information on either the session keys, input data from the client, or the result of the computation. Thus, the attacker gains information on the client data only if the computational leakage (during the Safe Computation phase in \Cref{fig:protocol}) reveals some information. It is therefore sufficient to show that the leakage is independent of the decrypted plaintext, in order to meet requirement \sreq.

\subsection{Safe computation}

The decrypted plaintext, shown in \Cref{fig:protocol}, is always marked as \blinded. Therefore, it is sufficient to prove that the safe computation functionality described in Section~\ref{sec:safe-comp} does not leak any information on any state marked as \blinded.

We do this by constructing a model \gls{ISA} using the F* programming language~\cite{fstar}, and proving that its visible and non-\blinded states are independent of the \blinded values in memory: that is to say, no changes to the \blinded values containing sensitive data, after a computational step, result in any change to the non-\blinded values or \blindedness bits that can be observed by the attacker.

Our approach is as follows:
\begin{enumerate}
    \item \label{safe-comp-step1}Define a low-level model of the server in terms of registers, memory cells, cache line allocations, and arbitrary transitions between states, and a security definition with respect to this model that meets the requirements of \Cref{sec:eval-protocol}.
    \item \label{safe-comp-step2}Define a more detailed architectural model that expresses instructions in terms of register and memory reads/writes, and a security definition that implies the low-level security definition from step \ref{safe-comp-step1}.
    \item Define a minimal instruction architecture that implements standard arithmetic operations with special-case taint propagation rules from \Cref{sec:taint-tracking}, and prove that it satisfies the security definitions from steps \ref{safe-comp-step1} and \ref{safe-comp-step2}.
\end{enumerate}

\newcommand{\equivstate}{\stackrel{\mathrm{state}}{\equiv}}
\newcommand{\equivlist}{\stackrel{\mathrm{list}}{\equiv}}
\subsubsection{Definitions}
We begin by defining equivalence relations on values that may be \blinded. We define a recursive type \prooflink{Multi.html}{\texttt{multiBlinded}}\footnote{We include links to specific modules and theorems where they are mentioned. The full model is available at \url{https://blinded-computation.github.io/blime-model/}.} that represents a blindable value:
\begin{minted}{FStar}
type blindedness_domain =
    x:FStar.UInt8.t{~(x = FStar.UInt8.zero)}
type multiBlinded (#t:Type) =
  | MultiClear: v:t 
      -> multiBlinded #t
  | MultiBlinded: v:t
      -> d:blindedness_domain
      -> multiBlinded #t
\end{minted}
Then, a \blinded value $v$ with \indicator $d$ is represented by the value $\texttt{MultiBlinded}~v~d$, and a non-\blinded value $v$ by $\texttt{MultiClear}~v$.

We say that two such values $a$ and $b$ are equivalent, denoted $a \equiv b$, if they are both \texttt{MultiBlinded} with the same \indicator, or if they are both \texttt{MultiClear} and have equal values.

We define equivalence of lists of \texttt{multiBlinded} values similarly: two lists $\ell_1$ and $\ell_2$ are equivalent, denoted $\ell_1 \equivlist \ell_2$, if their lengths are equal, and each of their values is equivalent.

\subsubsection{Low-level system model}\label{sec:fstar-low-level}
The system is modelled in \prooflink{Cpu.html\#system-state}{{Cpu.fst}} by a system state type containing the following data:
\begin{itemize}
    \item the program counter (\texttt{pc}), containing a 64-bit unsigned integer pointing into memory,
    \item an array of register values, each containing a blindable 64-bit unsigned integer,
    \item an array of memory values, each containing a blindable 64-bit unsigned integer, and
    \item an array of cache line assignments, each containing a 64-bit unsigned integer representing the address of the corresponding value in the cache.
\end{itemize}
We then define an equivalence relation $\equivstate$ (\prooflink{Cpu.html\#equivalence}{\texttt{equiv\_\0system}} in the model) on the system states $\mathcal{S}$, such that two system states are equivalent if they have equal program counters and cache line assignments, their registers and memory have equal \blindedness \indicators, and their non-\blinded values are equal.

We model the execution of instructions using a single-cycle fetch-execute model (\prooflink{Cpu.html\#execution-model}{\texttt{step}} in the model), with each instruction completing in a single cycle\footnote{To analyze the effects of features like speculation and variable-duration instructions, where instructions are not executed and committed sequentially, a more detailed microarchitecture-specific system model is needed where the model includes a wider range of internal processor state.}. An instruction $I \in \mathcal{I}$ is loaded from memory at the program counter address; if \texttt{pc} points to an instruction marked as \blinded, then it jumps to a fault handler at address $0$. Otherwise, the state of the processor is transformed according to an instruction-dependent execution mapping $X: \mathcal{I} \times \mathcal{S} \rightarrow \mathcal{S}$. We denote a single processor step
\[
   P_X(s) = \begin{cases}
                X(s.\textsc{memory}[s.\textsc{pc}], s), &\text{if} \; s.\textsc{pc} \; \text{points to} \\ &\text{non-\blinded memory,} \\
                s \text{ with s.\textsc{pc} = 0 } &\text{otherwise.}
          \end{cases}
\]

The security of the execution mapping $X$ is defined such that $X$ is secure if for all states $s_1$ and $s_2$, equivalent input states yield equivalent output states (\texttt{is\_safe} in the model):
\begin{align}
 \forall s_1, s_2 \in \mathcal{S}: s_1 \equivstate s_2 \Longrightarrow P_X(s_1) \equivstate P_X(s_2). \label{eqn:system-safety}
\end{align}
That is, $X$ is secure, if after each step in a computation, the values of \blinded registers and memory locations do not influence the \blindedness of any component of the output state, nor the values of any un\blinded value in the output, meaning that the attacker cannot infer anything about the \blinded state.

\subsubsection{Load-store model}\label{sec:fstar-load-store}
The low-level system model described in \Cref{sec:fstar-low-level} is simple and easy to understand, but since the execution mapping $X$ has unmediated access to the state, there is no easy way to express the taint propagation rule from \Cref{sec:design}. In \prooflink{InstructionDecoder.html}{InstructionDecoder.fst} we describe a higher-level model that allows us to better express statements about data flows between registers.

This model (the ``load-store model'') includes some microarchitectural details. Its execution mapping is defined in the function \prooflink{InstructionDecoder.html\#execution-model}{\texttt{decoding\_execution\_unit}} in terms of two functions:
\begin{itemize}
    \item \emph{An instruction decoder}, a function that takes as input an instruction word, and returns a decoded instruction containing an opcode, a list of input registers, and a list of output registers--either normal registers or \texttt{pc}.
    
    \item \emph{Instruction semantics}, a function that performs the actual computation, taking as input a decoded instruction and a list of \texttt{multiBlinded} register input values, and which returns a fault bit, a list of \texttt{multiBlinded} register output values with \blindedness bits, and a list of memory operations, each of which indicates a load or store between a register and an address in memory.
    
    \item \emph{A cache policy}, a function that accepts a set of cache line assignments and a memory operation, and returns a new set of cache line assignments.
\end{itemize}
The execution mapping then takes the instruction word, decodes it, reads the input operands from the initial system state, performs the computation, and if it does not raise a fault, increments \texttt{pc}, writes the results to registers, and performs the stores and loads, updating the cache line assignments.  In the event of a fault, \texttt{pc} is set to a fixed value.

The decoded instructions never depend upon \blinded data as, from \Cref{sec:fstar-low-level}, attempts to execute a \blinded value as an instruction results in a fault, and the execution mapping is never called.  Therefore, the safety of the execution mapping can be demonstrated by analyzing only the instruction semantics, as shown in the load-store model's \prooflink{InstructionDecoder.html\#main-safety-theorem}{main safety theorem}.

We define instruction semantics safety similarly to how we defined execution mapping safety in \Cref{eqn:system-safety}: for all instruction words, executing the instruction semantics function with equivalent lists of input operand values raises a fault in both cases, or neither case raises a fault and both yield equivalent output operand lists and memory operation lists\footnote{Memory operations are defined to be equivalent where they are between the same register and the same address in memory.}.

We then use this to prove the theorem \prooflink{InstructionDecoder.html\#main-safety-theorem}{\texttt{each\_\0load\0store\_\0exec\0ution\_\0unit\_\0with\_\0red\0acting\_\0equiv\0alent\_\0instr\0uc\0tion\_\0seman\0tics\_\0is\_\0safe}} that an execution mapping defined by any instruction decoder and safe instruction semantics is safe, no matter what cache policy is in use.

We next use the model to show the safety of a concrete \gls{ISA}.

\subsubsection{Model ISA}
Ideally, we would now apply the analysis above to a formal model extracted from the design of a real-world processor. However, this is a major undertaking in its own right that is beyond the scope of this paper.   We therefore analyze a simplified model \gls{ISA} with eight instructions in \prooflink{ISA.html}{ISA.fst}:
\texttt{STORE},
\texttt{LOAD},
\texttt{BZ} (branch if zero),
\texttt{ADD},
\texttt{SUB},
\texttt{MUL},
\texttt{AND}, and
\texttt{XOR}.

Each instruction accepts two input registers and one output register---the exceptions being \texttt{STORE} and \texttt{LOAD}, which require only two registers, one for the memory address, and one for the source or destination register respectively, and \texttt{BZ}, whose output is always written to \texttt{pc}.

Each instruction specifies the \blindedness of its outputs, as described in \Cref{sec:taint-tracking}. By default, each instruction marks its output as \blinded if any of its inputs are (consistently) \blinded, or raises a fault where some of its inputs are blinded but have inconsistent \blindedness \indicators.  However, there are some special cases that deviate from this treatment in order to better capture the data dependencies of the instructions and handle modifications of visible state:
\begin{itemize}
    \item \texttt{STORE} and \texttt{LOAD} instructions raise faults if their address input is blinded.
    \item \texttt{SUB} and \texttt{XOR} instructions yield an un\blinded value zero if both their inputs are the same register.
    \item \texttt{MUL} and \texttt{AND} instructions yield an un\blinded value zero if one of their inputs is an un\blinded value zero.
    \item \texttt{BZ} raises a fault if its comparison input is \blinded.
\end{itemize}

We then show in the model \gls{ISA}'s main safety theorem \prooflink{ISA.html\#safety}{\texttt{sample\_\0semantics\_\0are\_\0safe}} that these instruction semantics, defined in \texttt{sample\_semantics}, are safe as described in \Cref{sec:fstar-load-store}, and that the architecture with these semantics is therefore safe according to the definition given in \Cref{eqn:system-safety}.

This demonstrates that the taint tracking approach proposed in \Cref{sec:taint-tracking} is secure in general, and that the special cases that we have considered do not allow an attacker to violate the dataflow security definition from \Cref{eqn:system-safety}. This means that, so long as the external peripherals with which an outside observer can interact do not expose \blinded data, an observer cannot infer anything about the \blinded data in the system, satisfying objective \sreq.

\section{Performance \& resource usage evaluation}
\label{sec:perfEval}
\ifgemfive
\subsection{\implboom}
\fi
\label{sec:boomeval}

Incorporating \design into BOOM affects logical complexity (hence maximum clock rate), and the number of cycles that certain operations will take due to the additional memory accesses required to fetch the \blindedness \indicators. Therefore, we evaluate the overall performance of \implboom using both the maximum clock rate, and the number of clock cycles taken to execute a program.

We synthesized \implboom using the FireSim~\cite{firesim} FPGA-hosted simulation tool to measure the change in resource consumption and maximum clock rate. \Cref{tab:boompower} shows the power estimate and FPGA resource usage provided by Vivado, indicating low overheads for both \implboom-1 and \implboom-8.
Vivado timing analysis reported a Worst Negative Slack (WNS) of 0.018ns for both modified and unmodified cores, indicating no significant reduction to the maximum clock rate.

\begin{table}
    \centering %
    \begin{tabular}{r|r|rr|rr}
        & Unmod. & \multicolumn{2}{c|}{\implboom-1} & \multicolumn{2}{c}{\implboom-8} \\ \hline
        & Value & Value & $\Delta\;[\%]$ & Value & $\Delta\;[\%]$ \\
        \hline
        \textbf{Power (\si{\watt})} & $\mathbf{37.783}$ & $\mathbf{38.137}$ & $\mathbf{+0.9}$ & $\mathbf{38.319}$ & $\mathbf{+1.4}$ \\
        \hline
        \multicolumn{6}{c}{} \\
        \multicolumn{1}{c|}{\textbf{Resources}} & \multicolumn{5}{c}{} \\
        \hline
        \textbf{LUTs} & $\mathbf{460480}$ & $\mathbf{478950}$ & $\mathbf{+4.0}$ & $\mathbf{501847}$ & $\mathbf{+9.0}$ \\
        \hline
        \textbf{Registers} & $\mathbf{357179}$ & $\mathbf{371024}$ & $\mathbf{+3.9}$ & $\mathbf{388956}$ & $\mathbf{+9.0}$ \\
        \hline
    \end{tabular}  \vspace{1em}
    \caption{Effect of \implboom modifications on power consumption and FPGA resource usage (unmodified BOOM vs. \implboom-1 and \implboom-8).}
    \label{tab:boompower}
\end{table}

\begin{figure}
    \centering
    \includegraphics[width=0.95\linewidth]{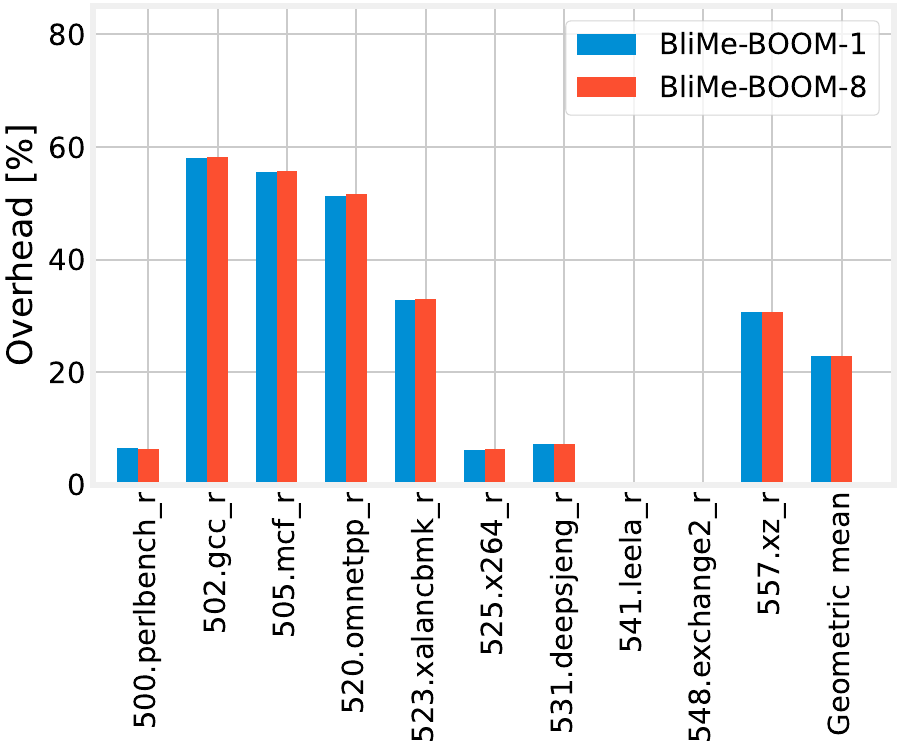}
    \caption{Overhead of \implboom-1 and \implboom-8 relative to unmodified BOOM, as measured using SPEC2017. The average overhead was \textbf{23\%} for both \implboom-1 and \implboom-8.}
    \label{fig:perfeval}
\end{figure}

To determine the effect of \implboom's memory modifications on performance, we ran the SPEC2017 Integer benchmark suite on both \implboom variations using the \inlinec{ref} workload. \implboom-1 was run with 8GiB of main memory, and \implboom-8 with 14GiB of main memory, so that both systems have 7GiB of addressable memory. This avoids the risk that the performance of the benchmarks is affected by differences in available memory. The results are shown in \Cref{fig:perfeval}; the average overhead across all benchmarks is \textbf{23\%} for both \implboom-1 and -8. This overhead, however, stems from the Chipyard limitation mentioned in \Cref{sec:boom}. As we show below in \Cref{sec:gem5}, appropriate memory optimization can substantially reduce this overhead.

\ifgemfive

\subsection{Memory optimization}\label{sec:gem5}
In \Cref{sec:boom}, we mentioned that an implementation optimized for small \blindedness \indicators can improve its performance by reducing the ratio of \blindedness \indicator requests to data requests. We investigate this by simulating both the optimized and unoptimized designs using the gem5 simulator~\cite{gem5}, where it is straightforward to reduce the memory request size for \indicators, and comparing their performance.

Gem5 is run in RISC-V full-system mode with the \inlinec{O3} out-of-order processor model. The cache configuration was chosen to match that used in \Cref{sec:boom}: 16kB each for private L1D and L1I, and 256kB for private L2. Main memory was 7GB of Dual Channel DDR3-1600 DRAM.  In addition, we modified the memory model to generate the extra requests used to read and write \blindedness \indicators, with the ratio of \blindedness \indicator to data request sizes being configurable as either 1:1 (as used in \implboom) or 1:8 (as in the proposed optimization).

Since gem5 is much slower than native execution using a \gls{FPGA}, following the approach taken in prior work~\cite{STT}, we ran the SPEC~CPU~2017 Integer benchmark suite with the \inlinec{ref} workload in each configuration and measured the number of cycles taken to execute and commit 1 billion instructions after skipping the first 10 billion instructions in order to allow the benchmark proper to start.  
\Cref{tab:boom-gem5} compares the performance results for \implboom-1, \implboom-8 and \design-gem5.
With the gem5 model configured in unoptimized mode, the average overhead of 25\% is similar to that of both \implboom-1 and -8 (23\%). 
With the model configured in optimized mode, the overhead reported by gem5 is reduced to \textbf{8\%}. The detailed benchmark results comparing the two configurations is shown in \Cref{fig:perfeval-gem5}. Therefore, we conclude that if the optimizations are implemented in hardware (by making the needed modifications to the memory subsystem in the Chipyard SoC as mentioned in \Cref{sec:boom}), similar overhead reductions can be achieved. As a result, for a \indicator-to-data size ratio of 1:8, 8\% is a fairer representation of \design{}’s overhead in real deployments by hardware manufacturers. Use of \indicator caches~\cite{efficientTagMem2017} would reduce overheads even further. The earlier figure of 23\%, on the other hand, corresponds to a \indicator-to-data size ratio of 1:1.

\begin{table}
    \centering %
    \begin{tabular}{r|r}
        \textbf{Implementation} & \multicolumn{1}{c}{\textbf{Avg. overhead (\%)}} \\
        \hline
        \implboom-1 & $\mathbf{23}$ \\
        \implboom-8 & $\mathbf{23}$ \\
        \design-gem5 & $\mathbf{25}$ \\
        \design-gem5 Optimized & $\mathbf{8}$ \\
    \end{tabular}  \vspace{1em}
    \caption{Average overheads of running SPEC2017 on \implboom-1 and -8, \design-gem5 and \design-gem5 Optimized. The averages are calculated for the full benchmarks on \implboom-1 and -8, and for 1 billion instructions on \design-gem5 and \design-gem5 Optimized. The gem5 simulation of the unoptimized system shows an average overhead of 25\%, similar to that of both \implboom-1 and -8 (23\%).}
    \label{tab:boom-gem5}
\end{table}

\begin{figure}
    \centering
    \includegraphics[width=0.95\linewidth]{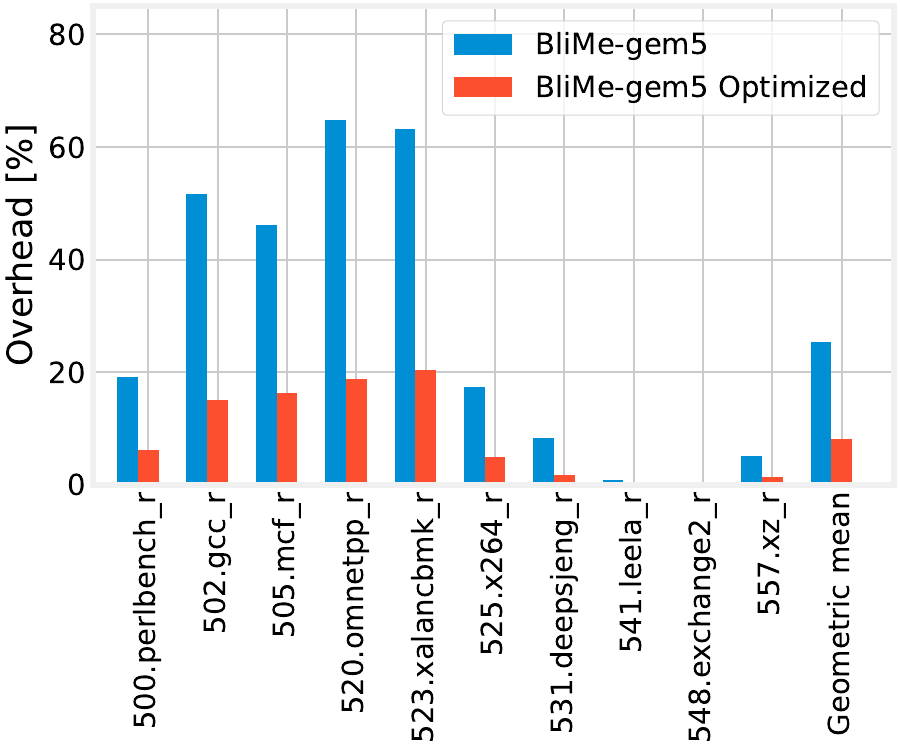}
    \caption{Overhead before and after applying the optimization from \Cref{sec:boom}, as measured using gem5 simulator to execute 1 billion instructions of SPEC2017. The optimization reduces the average overhead on gem5 from 25\% to 8\%.}
    \label{fig:perfeval-gem5}
\end{figure}

\fi

The closest comparisons to \design, \gls{OISA}~\cite{yu19} and \gls{FHE}, suffer from significantly higher overheads. \gls{OISA}'s fastest benchmarks (matrix-mult and neural-net) incur approximately 35\% overhead with larger-than-cache inputs, whereas \gls{FHE} is several orders of magnitude slower~\cite{viand21}. Overall, given the moderate impacts on performance and resource usage discussed in \Cref{sec:boomeval,sec:gem5}, we %
conclude that requirement \preq is satisfied.

\section{Compatibility evaluation}%
\label{sec:compEval}

For backwards compatibility with existing code (requirement \creq), two types of code are important: code that processes exclusively non-\blinded data, and code that is already side-channel resistant and processes \blinded data.

\design's taint tracking policy (\Cref{sec:taint-tracking}) changes the behavior of the processor only where an instruction depends upon a \blinded input value. Therefore, code that exclusively processes non-\blinded data is unaffected.

Code that processes \blinded data will fault if it attempts to modify an observable output in a way that the processor determines to be dependent on the \blinded data. To evaluate whether the \design implementations have met requirement \creq, we must therefore determine whether side-channel-resistant code will comply with the \design taint-tracking policy.

\design faults when \blinded data attempts to flow to non-\blindable observable outputs (\Cref{sec:taint-tracking}).
Existing side-channel-resistant code already prevents the program counter and the addresses of memory operations from being affected by \blinded data, e.g., as in~\cite{bernstein12}. Data flows of \blinded data to other visible state, execution time and fault signals (\Cref{sec:taint-tracking}) are prevented by design. Therefore, side-channel-resistant code is compatible with \design's taint-tracking policy, so long as only sensitive data is \blinded. We successfully ran stream cipher encryption/decryption with a \blinded key using the TweetNaCl~\cite{tweetNaCl} library, demonstrating the backwards compatibility of \design.

However, the \gls{CPU} cannot identify cases where the result of a computation remains \blinded even though it no longer contains any sensitive data. For example, it will not be possible to branch on the result of the decryption/verification of an authenticated encryption, meaning that its cryptographic \glspl{API} will need to be modified so that its control flow does not depend on whether the verification was successful. In practice, this means that any computation must always continue as though verification was successful, with any failure being indicated by a flag that is returned to the client (\Cref{sec:secret-dep-faults}).

\section{Future work}\label{sec:discussion}

\subsection{Compiler support to improve deployability}%
\label{sec:compiler}

The security guarantees of \design are enforced solely by the hardware modifications without requiring any compiler support. We saw that some existing side-channel resistant code successfully runs on \design (\Cref{sec:compEval}). Developers can write new \design-compliant code. However, this can be a challenge, particularly for larger programs because 
1) manual data-flow tracking can be too complex, and 
2) \design-unaware compilers might inadvertently induce data-flow dependencies not evident in the source language.
Consequently, compiler support can increase deployability by (a) \emph{verifying} code compatibility, and where possible (b) automatically \emph{transforming} non-compliant code to an equivalent compliant variant, or (c) semi-automatically transforming the code by identifying problematic pieces of code to be manually rewritten by the developer.

To verify compliance, one approach is to use binary analysis. This has a number of drawbacks. Binary lifting is itself error-prone~\cite{kimTesting2017}, and could induce uncertainty. Focusing on binary representation makes it difficult to generate useful error messages. Instead, we opt for augmenting the compiler so that we can use the \gls{IR} for transformations, provide useful error messages through the front-end, and control the machine-code generation.

To transform code, we can employ known rules for safe coding like \emph{cryptocoding}~\cite{aumassonCryptocoding2022}, and use compiler-specific approaches that aid both analysis and verification.
Specifically, we can use techniques such as \emph{function cloning} to improve the accuracy of the static data-flow tracking and to avoid instrumentation of data-flows that at run-time are not necessarily \blinded.
For instance, a function could be called in multiple places, both such that its arguments are never tainted, and that they may be tainted.
By cloning the function we can avoid needlessly instrumenting and tainting the non-\blinded data-flow and instead only taint the cloned variant of the function.
We are currently exploring both the adaption of existing software-only approaches such as Constantine~\cite{borrelloConstantine2021}, and implementing \design-specific analysis and transformations.

Because complete program analysis is undecidable~\cite{riceClasses1953}, we cannot guarantee that any approach detects all compliant code.
Given memory-safe code, we require a sound analysis that rejects all non-compliant code and accepts only code that will not result in a taint-tracking policy violation fault at run-time.

\subsection{Handling large numbers of clients}\label{sec:discussion:trustedOS}
\design as described above can handle $2^n - 1$ simultaneous clients with an $n$-bit blindedness \indicator.  This means that the ability to handle larger numbers of clients will require a logarithmic increase in memory overhead. This can be made configurable at boot-time, allowing servers to trade main memory capacity for a larger number of simultaneous clients. Another approach, however, is to overcome this overhead by multiplexing the same \blindedness \indicator value across multiple clients, and using an \gls{OS} trusted by clients---e.g., by using remote attestation of a well-known certified \gls{OS}---to prevent data from flowing between programs that share \blindedness \indicator values.  In the extreme case, this reduces to a single-bit blindedness \indicator, as implemented in \implboom-1.

This \gls{OS} has several tasks. On context switch, the \gls{OS} asks the encryption engine to export the current client key by ``sealing'' it (i.e., encrypting it using a key known only to the encryption engine), and provides a previously sealed key, corresponding to the incoming process, which the encryption engine can unseal and set as the new current client key. Thus, even though the \gls{OS} never sees any client key in the clear, it must be trusted to load the correct sealed key onto the encryption engine.

The normal process isolation features of the \gls{OS} ensure that a client cannot access another client's \blinded data directly or any of the client session keys held by the \gls{OS}. Because different processes may share \blindedness \indicator values, the \gls{OS} must further ensure that an application cannot transfer \blinded data to other applications via system calls or other \gls{IPC} mechanisms.
Preventing a client from accessing another client's \blinded data ensures two things: 1) an application serving one client cannot use the encryption engine to encrypt and \unblind \blinded  data originating from another client, and 2) computation can never use \blinded data belonging to two different clients, since there would be no clear way to determine the ownership of the result. Therefore, processed data is only encrypted and exported with a key corresponding to the client from which the input data was originally imported. Note that we only rely on the OS to prevent clients from having \emph{direct} access to other clients' \blinded data. We do not rely on the OS code being side-channel resistant, because \blinded data is protected from side channels by the hardware.

\subsection{Enabling safe local processing}
\design extensions can be usefully applied on the local machine as well. Since an application processing \blinded data cannot infer anything about the data other than its length, it can safely process data belonging to other users or applications.

For example, the \gls{OS} can allow an application to read data from a file that is normally inaccessible to the application with the constraint that any data read will be marked as \blinded. This makes it possible to build useful computational pipelines while strongly adhering to the principle of least privilege.

\subsection{Handling secret-dependent faults}
\label{sec:secret-dep-faults}
In \design, a fault's occurrence cannot depend on a secret value, since the fact that it occurs (or not) can leak information. For example, if a div-by-zero fault will occur if the divisor is a \blinded data item with a value zero, and this faults leads to an interrupt handler being called, the change in control flow will reveal that the \blinded value was zero. On the other hand, if it does not occur, it reveals that the value was not zero.  We therefore do not allow \blinded values to be used in such situations.
To avoid this limitation, we might instead suppress faults depending on \blinded data, so that the control flow remains \emph{as if the fault did not occur}. The client can be informed of this fault and that the computation results are invalid by setting a \blinded bit in some protected storage (e.g., a special register) when the fault occurs, so that the software can convey this bit to the client as part of the returned encrypted results. 

\subsection{Defending against other attacks}
\label{sec:discussion:other-attacks}
Rowhammer~\cite{originalRowhammer} is a vulnerability in modern \gls{DRAM} modules that threatens the integrity of data. Due to the high proximity of DRAM cells, toggling a row of cells at a sufficiently high rate can result in bit flips in adjacent rows. Exploits of this phenomenon by a remote adversary have been continuously demonstrated despite the number of proposed defenses~\cite{rowhammerRetro}. %
We make the common and reasonable assumption that reading data from any address in memory retrieves the last value that was written to that address; this includes the extra \blindedness \indicator. Rowhammer is an orthogonal vulnerability requiring orthogonal defenses to ensure memory integrity. Other fault-injection attacks based on \gls{DVFS}, such as CLKscrew~\cite{clkscrew}, V0LTpwn~\cite{voltpwn} and Plundervolt~\cite{plundervolt}, produce similar attack patterns and are also out of scope.

Attacks that require physical access to the server are also out of scope. Physical access enables full read-write side-band access to memory through direct connection to the \gls{DRAM} bus. It also facilitates more powerful side-channel attacks, such as those relying on power~\cite{kocher99dpa}, electromagnetic~\cite{emAttack1,emAttack2} or temperature measurements~\cite{tempAttack}, or fault-injection attacks, such as VoltPillager~\cite{voltpillager}, that can break \gls{HSM} confidentiality and integrity guarantees. We leave an adversary model that includes physical access to future work.

\section{Related work}
\label{sec:related}

\textbf{Taint tracking.}
A large body of work exists on taint tracking, also called \gls{DIFT} \cite{dynamicIFT2004}. Hu \emph{et al}.~\cite{huHardwareInformationFlow2021} present a survey that includes several hardware-based taint-tracking techniques with varying goals and security/performance trade-offs, and at different abstraction layers. Speculative Taint Tracking~\cite{STT} applies taint tracking to the results of speculatively executed instructions to prevent them leaking information. Tiwari \emph{et al}.~\cite{GLIFT} propose taint tracking at the gate level, and use it to create a processor that is able to track all information flows, but has a limited \gls{ISA} and suffers from large overheads.
Taint tracking can also be performed purely in software. Data flow integrity is a form of taint tracking that protects software against non-control data attacks by using reaching definitions analysis~\cite{dfi}. Pointer tainting~\cite{chen2005a} is another defense against non-control data attacks that taints user input and detects an attack when a tainted value is dereferenced as a pointer.

\textbf{Data-oblivious execution.}
Preventing side-channel leakage requires covering several observable outputs. %
Several algorithms have been proposed to obfuscate data-dependent memory access patterns~\cite{ORAM,pathORAM2018,permuteRAM2017,optorama2020}. However, they all come at a significant cost to performance. Other work has focused on making code constant-time to prevent leakage through execution timing. %
An example is Constantine~\cite{borrelloConstantine2021}, which extends LLVM to compile code into constant-time binaries. It relies, however, on dynamic analysis to identify vulnerable code for transformation. This can be imprecise as full execution path coverage is not guaranteed, potentially leading to some vulnerable code not being transformed.

Lee \emph{et al}.~\cite{dove} propose DOVE to protect sensitive data used in outsourced computation from side channels. DOVE uses a frontend to transform the client application code to a custom data-oblivious representation called a Data Oblivious Transcript (DOT). The DOT is then sent to a trusted interpreter (the backend) on the server, which verifies that the DOT is data oblivious and then runs it on the sensitive data. The trusted interpreter must be run within a \gls{TEE}, such as an Intel SGX enclave, as it is part of the \gls{TCB}.

Yu \emph{et al}.~\cite{yu19} develop an \gls{OISA} that performs run-time taint tracking of sensitive values and adds a duplicate set of instructions to the \gls{ISA}. Each operand of the additional instructions is defined as either safe or unsafe. Using any tainted values as unsafe operands results in a fault. The hardware guarantees that computation is oblivious to safe operands and, therefore, that any sensitive values used as safe operands are not leaked through side channels. %
\gls{OISA} offers taint and untaint instructions and relies on the application code to use them correctly to taint/untaint sensitive values during computation. Consequently \gls{OISA} is \emph{not applicable for our usage scenario} described in \Cref{sec:usage}, where outsourced computation is carried out by potentially \emph{untrusted, third-party}, application code, for multiple simultaneous clients. Although \gls{OISA} could use remote attestation to verify the server-side application code, remote attestation of arbitrary third-party application code is neither realistic nor scalable.  Furthermore, software vulnerabilities in the application code can allow adversaries to untaint arbitrary sensitive values through the \gls{OISA}'s untainting instruction, which is exposed to any software running on the main CPU.

\textbf{Point solutions for side-channel attacks.}
The literature contains a variety of side-channel attacks that leak information through the processor's caches~\cite{flushReload,flushFlush,primeProbe,gruss2015,gruss2016}. Defenses against these attacks rely on temporal or spatial isolation between processes; the cache is either flushed on context switches or is partitioned in such a way that each process uses a separate fixed portion of the cache~\cite{partitionedCache,PLCacheRPCache}. However, this results in unnecessary overhead when processing non-sensitive data. Other methods change the cache architecture or replacement policy but also suffer from unnecessary overheads~\cite{zcache,secRandCache}.
The cache attacks mentioned above are usually used as building blocks to create covert channels for more sophisticated attacks such as Meltdown and Spectre~\cite{Lipp2018meltdown,Kocher2018spectre}. In response, a range of defenses have been proposed to stop these sophisticated attacks~\cite{STT,kaiser}. However, they do not address the main source of information leakage (which is the data-dependent memory access pattern) but rather provide point solutions for specific attacks.%

\section{Conclusion}
We introduced \design, a new approach to outsourced computation that uses hardware extensions to ensure that clients' sensitive data is not leaked from the system, even if an attacker is able to run malware, exploit software vulnerabilities, or use side channel attacks. \design does this while maintaining compatibility with existing side-channel-resistant code, and without reducing performance significantly. In designing \design, we follow the design pattern of using a separate, discrete hardware security component in conjunction with the main \gls{CPU}, common on both servers~\cite{tpm2lib} and end user devices~\cite{titanM}.
By using such a remotely attestable fixed-function \gls{HSM} in combination with taint-tracking \gls{ISA} extensions, \design can provide functionality similar to that of fully homomorphic encryption, but achieving native-level performance by replacing cryptography with hardware enforcement.

\section*{Acknowledgements}
This work is supported in part by Natural Sciences and Engineering Research Council of Canada (grant number RGPIN-2020-04744), Intel (in the context of the Private AI consortium), the Government of Ontario, and the Academy of Finland (decision 339514). Views expressed in the paper are those of the authors and do not necessarily reflect the position of the funders. Our performance evaluation was facilitated in part by a gift of cloud credits from Amazon Web Services.

\begin{appendices}

\section{Examples of \design-compliant applications}

In this section we present two examples of code that adheres to constant-time principles and can therefore run without faulting on \design. We also show how to use \design's blinding and unblinding operations with these examples. Both examples were adapted from the OISA work by Yu \emph{et al}.~\cite{yu19}. The first example is data-oblivious matrix multiplication and requires no changes. The second is a function to find the maximum value in an array. We present the initial code and then show how to manually transform it to make it data-oblivious.

\subsection{Matrix multiplication}

\Cref{lst:matrixmult} shows the code for matrix multiplication. The code is already data-oblivious, i.e., adheres to the restrictions in \Cref{sec:compliant-software}. It therefore does not require any changes. In \Cref{lst:matrixmult}, we also show how encrypted data from the client is preprocessed to decrypt-and-blind it, and how the result is postprocessed to encrypt-and-unblind it.

\begin{listing}
\begin{minted}%
{C}

// int *A, *B, *C;
// with sizes szA, szB, szC, respectively

// Here, the input matrices A and B are
// populated with unblinded encrypted data 
// from the client, e.g., through some
// network interface

// We first decrypt-and-blind A and B:
blind(A, szA);
blind(B, szB);

// Now, A and B contain blinded plaintext

int C_nrow = A_nrow;
int C_ncol = B_ncol;
assert (A_ncol == B_nrow);
for(int i = 0; i < C_nrow; i++){
  for(int j = 0; j < C_ncol; j++){
    for(int k = 0; k < A_ncol; k++){
      C[i*C_ncol + j] += 
          A[i * A_ncol + k] * B[k * B_ncol + j];
    }
  }
}

// Now, C contains the blinded plaintext result
// We must then unblind C ...

unblind(C, szC);

// ... which gives us the result as
// unblinded ciphertext that we can then
// send back to the client

\end{minted}
\caption{Data-oblivious code for matrix multiplication.}\label{lst:matrixmult}
\end{listing}

\subsection{Finding the maximum}

\Cref{lst:findmax1} shows non-data-oblivious code for finding the maximum value in an array of integers. The blinding and unblinding operations are similar to those in \Cref{lst:matrixmult}, and we therefore skip them for brevity.

Initially, the only \blinded data is that inside \texttt{arr}. In the first iteration, \blinded data from \texttt{arr} gets loaded into \texttt{max\_val} on line 6. In the second iteration, the \blinded value in \texttt{max\_val} is used in a branching condition on line 4, resulting in a \design violation and a fault.

We can make this code data-oblivious by linearizing the if condition using predicated execution. The transformed code is shown in \Cref{lst:findmax2}.

\begin{listing}
\begin{minted}%
{C}
int FindMax(int arr[] /*blinded*/, 
            int N     /*unblinded*/){
  int max_val = -1;
  for (int i = 0; i < N; i++){
    if (arr[i] > max_val){
      max_val = arr[i];
    }
  }
}
\end{minted}
\caption{Non-data-oblivious code for finding the maximum value in an array of integers.}\label{lst:findmax1}
\end{listing}

\begin{listing}
\begin{minted}%
{C}
int FindMax(int arr[] /*blinded*/, 
            int N     /*unblinded*/){
  int max_val = -1;
  for (int i = 0; i < N; i++){
    int predicate = arr[i] > max_val;
    max_val = (predicate * arr[i]) | 
                (!predicate * max_val);
  }
}
\end{minted}
\caption{Data-oblivious code for finding the maximum value in an array of integers after manual transformation.}\label{lst:findmax2}
\end{listing}

\section{Artifact Appendix}

\subsection{Description \& Requirements}

\design's evaluation consists of three parts: security (Section VI), performance (Section VII - Table III, Figures 4) and resource usage (Section VII - Table II).

\begin{itemize}
    \item \textbf{Security}: The security evaluation is done by running the provided model in F* and confirming that it passes verification.
    \item \textbf{Cycle-accurate performance}: The performance evaluation is done by running SPEC CPU17 on two gem5 implementations, \design-gem5 and -gem5 Optimized. These can be run entirely on commodity hardware but are only used to run 11 billion instructions of each SPEC CPU17 benchmark; running the benchmarks to completion requires an infeasible amount of time. The performance of \design-gem5 is an estimate of the performance of \implboom-1 and -8. This is because the only performance overhead introduced by \design is the additional memory accesses to tags, which take the same number of cycles for both \implboom-1 and \implboom-8. \design-gem5 Optimized corresponds to an optimized version that is not possible on \implboom due to a limitation in Chipyard not \design (Section V.B.2).
    \item \textbf{Resource usage \& clock frequency}: The power, resource usage and clock frequency overheads are obtained by synthesizing \implboom-1 and -8 using Xilinx Vivado on an \gls{AWS} instance. This is done by running FireSim scripts.
\end{itemize}

\subsubsection{How to access}

Our artifact is available at \artifacturl, and archived at \url{https://doi.org/10.5281/zenodo.10161487}.

\subsubsection{Hardware dependencies}

Obtaining the overheads requires the use of \gls{AWS}. This is because we use FireSim scripts that must be run on \gls{AWS}.

\subsubsection{Software dependencies} 

F* and gem5 implementations were tested on Ubuntu 20.04.

\subsubsection{Benchmarks} 

We use the SPEC CPU 2017 benchmark suite for the gem5 performance evaluation, which must be obtained separately from \url{https://www.spec.org/cpu2017/}.

\subsection{Artifact Installation \& Configuration}

Docker is required for security evaluation using F*.  Installation then proceeds according to the README.

The full FireSim evaluation requires an AWS instance running the AWS FPGA Developer AMI. The README in the repository describes how to start and configure a suitable instance.  Installation then proceeds according to the README.

The gem5 evaluation requires a Ubuntu Linux system.  Installation then proceeds according to the README in the repository.

\subsection{Major Claims}

\begin{itemize}

    \item (C1): A simplified model \gls{ISA} with BliMe extensions is formally verified to maintain the confidentiality of \blinded data.

    \item (C2): An unoptimized \design implementation incurs a 25\% performance overhead. Optimization reduces this to 8\%.

    \item (C3): \design has low power and FPGA resource usage overheads.

\end{itemize}

\subsection{Evaluation}\label{sec:eval}

\subsubsection{Experiment (E1)}
[1 human-minute + 1--2 compute-hours]: This machine-checks the F* model to verify that the confidentiality of \blinded data cannot be broken (C1).

\textit{[Preparation]}
Build the Docker container from the \texttt{model/} directory of the repository.

\textit{[Execution]}
Run the Docker container built in \emph{[Preparation]} without arguments.

\textit{[Results]}
The container will output the message \texttt{All verification conditions discharged successfully}.

\subsubsection{Experiment (E2)}\label{E2}
[30 human-minutes + 24 compute-hours]: This runs 11 billion instructions of each SPEC CPU 2017 benchmark on \design-gem5 and \design-gem5 Optimized. The first 10 billion instructions are used to warm-up the system. The results of E2 provide an estimate for claim C2 in terms of number of cycles. We show in E3 that the time for each cycle (clock frequency) does not change, which means that the performance overhead is dependent \emph{only} on the change in the \emph{number} of cycles, i.e., E2.

\textit{[Preparation]} 
Follow the instructions in the \texttt{gem5} folder's README file.

\textit{[Execution]}
Follow the instructions in the README file. This will run the 10 SPEC CPU 2017 Integer benchmarks on the unmodified gem5, \design-gem5 and \design-gem5 Optimized. Two \texttt{tmux} sessions will be spawned per benchmark: one (with \texttt{\_run\_} in its name) to run the gem5 simulation and one (with \texttt{\_telnet\_}) to \texttt{telnet} into the gem5 simulation and run the benchmark. Once the benchmark is complete, the gem5 simulation will exit. This can be checked by attaching to the \texttt{\_run\_} tmux session using \texttt{tmux a -t <name-of-tmux-session>}.

\textbf{Note:} Benchmarks can rarely crash prematurely. We have experienced this on both modified and unmodified gem5. Rerunning the benchmark should fix the issue. Please refer to the README-single file for information on how to rerun a single benchmark.

\textit{[Results]}
Once all the gem5 simulation have exited, the results will be in the \texttt{m5out/stats.txt} files in each benchmark directory. Follow the instructions at the end of the README file to extract the \gls{IPC} numbers.

\subsubsection{Experiment (E3)}
[30 human-minutes + 18 compute-hours]: This synthesizes \implboom-1 and -8 using FireSim on \gls{AWS}. The synthesis will produce power and resource usage reports that support claim C3. It will also produce a timing report that shows no reduction in maximum clock frequency, supporting our argument in \Cref{E2}.

\textit{[Preparation]}
Follow the instructions in the \texttt{firesim} folder's README file to set up the AWS EC2 instance.

\textit{[Execution]}
Follow the instructions in the README file. This will clone our modified FireSim repository and run the scripts to build the FPGA bitstream.   NB: Building the bitstream will use several hundred US Dollars' worth of computation.  Following the README's additional instructions to run the full SPEC 2017 benchmark will cost approximately US\$5000.

\textit{[Results]}
The power, timing and resource usage reports will be available in \texttt{firesim/deploy/results-build/\\<timestamp-config>/<design>/build/reports/}. Power \& resource usage results can be found in \texttt{<timestamp>.SH\_CL\_final\_power.rpt} under ``Total On-Chip Power'' \& ``On-Chip Components''. Clock frequency results can be found in \texttt{<timestamp>.SH\_CL\_final\_timing\_summary.rpt} under ``WNS(ns)''.

\subsection{Known Issues}
\begin{itemize}
    \item Attempting to recursively clone submodules for the FireSim experiments can result in an error about unreachable submodules. Please do not attempt to handle submodules directly. The build scripts used in the README instructions will fetch the submodules correctly.
\end{itemize}

\end{appendices}

\bibliography{refs}

\end{document}